\documentclass[12pt]{article}

\usepackage{amsmath, amssymb, amsfonts}
\usepackage{geometry}
\usepackage{graphicx}
\usepackage[most]{tcolorbox}
\usepackage[colorlinks=true, allcolors=blue]{hyperref}
\usepackage{cite}
\usepackage{authblk}
\usepackage{verbatim} 
\usepackage{tikz} 
\usetikzlibrary{arrows.meta, bending, positioning, shapes.geometric}

\newcommand{\bra}[1]{\left\langle #1 \right|}
\newcommand{\ket}[1]{\left| #1 \right\rangle}
\newcommand{\braket}[2]{\left\langle #1 \middle| #2 \right\rangle}
\newcommand{\matrixel}[3]{\left\langle #1 \middle| #2 \middle| #3 \right\rangle}

\begin{document}

\title{\textbf{Weak Values as Geometric Lenses: \\ Deformations of Hilbert Space and the Emergence of superoscillations}}
\author{Mirco A. Mannucci}
\affil{Holomathics, LLC}
\date{July 20, 2025}
\maketitle

\begin{abstract}
The formalism of weak measurement in quantum mechanics has revealed profound connections between measurement theory, quantum foundations, and signal processing. In this paper, we develop a pointer-free derivation of superoscillations, demonstrating that they are a natural and necessary consequence of the geometric structure underlying weak values. We argue that the weak value is best understood as a \textbf{ratio of geometric deformation}, quantifying how an observable transforms the structure of Hilbert space relative to a reference provided by the standard inner product. This deformation acts as a conceptual lens, warping the local structure of quantum states to produce oscillations far exceeding the global Fourier bandwidth. We formalize this by interpreting the weak value as a comparison between a deformed sesquilinear form and the standard one, and explore its deep connections to generalized Rayleigh quotients and the projective geometry of quantum states. This perspective unifies weak values and superoscillations as two facets of a single underlying geometric principle.
\end{abstract}


\tableofcontents
\newpage

\section{Introduction}
This paper, which continues the investigation into the algebro-geometric notion of weak values begun in\cite{Mannucci2025},  introduces a new perspective by framing the phenomenon of superoscillation \footnote{A superoscillation is a locally high-frequency component of a function that globally remains bandlimited.} not as a measurement outcome, but as an intrinsic feature of a deformed Hilbert space geometry. 

We show that superoscillations are a necessary consequence of this geometric warping. The article is structured to build this geometric argument from the ground up: Section 2 introduces the core concept of the weak value as a ratio of deformation. Section 3 explores its deeper mathematical connections. Section 4 provides a concrete visualization of this deformation on the Bloch sphere. Section 5 explains how this geometric lens effect is the direct cause of superoscillations. Finally, the remaining sections offer a conclusion and future outlook.

\section{The Weak Value as a Ratio of Geometric Deformation}
The geometric essence of the weak value is revealed by analyzing its structure. For a pre-selected state $\ket{\psi}$, a post-selected state $\ket{\phi}$, and an observable $A$, the weak value is:
\begin{equation}
    A_w = \frac{\matrixel{\phi}{A}{\psi}}{\braket{\phi}{\psi}}
\end{equation}
This is not merely a formula for a measurement outcome; it is a statement about the geometry of Hilbert space. The denominator, $\braket{\phi}{\psi}$, is the standard inner product, establishing the baseline geometric relationship. The numerator, $\matrixel{\phi}{A}{\psi}$, represents a deformation of this geometry. The weak value $A_w$ is the ratio that quantifies this deformation. When the post-selection is nearly orthogonal to the pre-selection ($\braket{\phi}{\psi} \to 0$), the geometry is sharply magnified, and even a mild deformation becomes a large effect.

\section{Further Mathematical Structures and Generalizations}

\subsection{The Weak Value as a Generalized Rayleigh Quotient}
The expectation value of an observable $A$ in a pure state $\ket{\psi}$ is given by the classic Rayleigh quotient. The weak value generalizes this structure by utilizing two distinct directions:
\begin{equation}
    A_w = \frac{\matrixel{\phi}{A}{\psi}}{\braket{\phi}{\psi}},
\end{equation}
which can be understood as a \textbf{"bi-orthogonal" Rayleigh quotient}. In this context, $A_w$ can assume values well outside the spectral range because the numerator is a sesquilinear form that is not positive-definite, and the denominator can approach zero arbitrarily without divergence of the numerator \cite{Dressel2014}. From a functional analysis perspective, $A_w$ can be interpreted as a \emph{complex directional derivative} of the representation of $A$ along trajectories in the tautological bundle over $\mathbb{CP}(\mathcal{H})$.That is, $A_w$ measures how rapidly the expectation of $A$ changes as the state moves infinitesimally along a path in projective Hilbert space—a geometric interpretation of sensitivity to deformation. \footnote{The \emph{tautological bundle} over projective Hilbert space $\mathbb{CP}(\mathcal{H})$ is the bundle whose fiber over each ray $[\psi]$ consists of the line spanned by $\ket{\psi}$ itself. It provides a natural setting to describe lifted quantum trajectories in Hilbert space, making directional derivatives geometrically meaningful.}

\subsection{Projective Geometry and Meromorphic Structure}
Physical states correspond to points in a complex projective space, $\mathbb{CP}(\mathcal{H})$. The weak value is a \textbf{rational} (or \textbf{meromorphic}) function on the product space $\mathbb{CP}(\mathcal{H}) \times \mathbb{CP}(\mathcal{H})$. Its poles occur when $\braket{\phi}{\psi} = 0$. The phenomenology of large weak values is the study of this function in the neighborhood of its polar divisor. The mathematical properties of super-oscillatory functions have been studied extensively \cite{Colombo2011, Sabadini2019}.

\section{Observable-Induced Geometry on the Bloch Sphere}
To build intuition, let us start with the simplest non trivial quantum system, the qubit, and its standard geometrical representation, the Bloch sphere. 

To visualize weak values as local distortions of geometry, consider a qubit with observable $A = \vec{a} \cdot \vec{\sigma}$. A pure qubit state $\ket{\psi(\theta, \phi)}$ corresponds to a point on the Bloch sphere:
\[
\ket{\psi(\theta, \phi)} = \cos(\theta/2)\ket{0} + e^{i\phi} \sin(\theta/2)\ket{1}.
\]
The standard metric on the sphere is the \textbf{Fubini–Study metric}: $ds^2 = \frac{1}{4}(d\theta^2 + \sin^2\theta \, d\phi^2)$. Now replace the standard inner product with the operator-induced sesquilinear form:
\[g_A(\psi, \psi) = \braket{\psi | A | \psi} = \vec{a} \cdot \vec{n}(\theta, \phi).\]
This defines a scalar field over the sphere, and its differential defines a new metric:
\[
ds_A^2 = \left( \frac{\partial g_A}{\partial \theta} \right)^2 d\theta^2 + 2 \frac{\partial g_A}{\partial \theta} \frac{\partial g_A}{\partial \phi} d\theta d\phi + \left( \frac{\partial g_A}{\partial \phi} \right)^2 d\phi^2.
\]
The result is an anisotropic geometry—a local deformation of distances and angles depending on $A$. The weak value compares this geometry to the standard one from the viewpoint of a fixed $\bra{\phi}$.

\begin{figure}[h!]
\centering
\begin{tikzpicture}[scale=2]
    \shade[ball color=gray!10, opacity=0.4] (0,0) circle (1);
    \draw[gray!60] (0,0) circle (1);
    \draw[gray!50, dashed] (0,1) arc (90:270:0.3 and 1);
    \draw[gray!50] (0,1) arc (90:-90:0.3 and 1);
    
    \draw[black, thick, dashed, rotate around={30:(0,0)}] (0,0) ellipse [x radius=1.2, y radius=0.8];
    
    \draw[->, thick, blue!60!black] (0,0) -- (0.5, 0.866*0.8) node[above right] {$\vec{a}$};

    \node[black, align=center] at (1.1, 0.9) {Ellipsoidal geometry\\induced by $A=\vec{a}\cdot\vec{\sigma}$};
\end{tikzpicture}
\caption{Illustration of metric deformation on the Bloch sphere: a local ellipsoidal distortion induced by the observable $A = \vec{a} \cdot \vec{\sigma}$.}
\label{fig:bloch_deformation}
\end{figure}
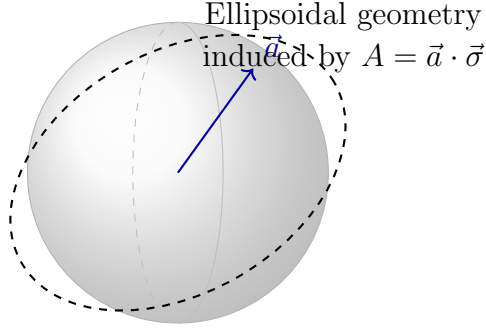

\begin{tcolorbox}[colback=gray!3, colframe=black, title=An Intuitive Picture of Weak Values]
Imagine you are standing in a landscape. You see two trees—one in the foreground, one in the distance—and you fix your gaze between them. The distance between them defines a baseline: your ordinary perception of geometry.

Now suppose you put on a strange pair of glasses—glasses that don’t just filter light, but subtly warp the geometry of space itself. These glasses correspond to a quantum observable \( A \). As you look again at the two trees, you perceive their relative distance as altered—perhaps elongated, compressed, or even twisted in phase.

The \textbf{weak value} is the ratio between what you now perceive (with the warped geometry) and what you perceived before. It tells you: \emph{how much has this observable deformed my view of the world between these two fixed reference points?}

Just as lenses can magnify or distort space, observables can distort Hilbert space. The weak value measures that distortion—a geometric magnification factor that can grow arbitrarily large near special alignments. This is why weak values are not just strange numbers—they are the quantum version of perspective itself.
\end{tcolorbox}

\section{Geometry of superoscillations: From Weak Values to Local Frequency Amplification}
Consider the unitary evolution operator $U(\theta) = e^{i\theta A}$ with $A$ Hermitian and spectrum in $[-\Lambda, \Lambda]$. Let $\ket{\psi}$ and $\ket{\phi}$ be nearly orthogonal. We define the weak evolution amplitude:
\begin{equation}
F(\theta) := \frac{\braket{\phi}{e^{i\theta A} \psi}}{\braket{\phi}{\psi}}.
\end{equation}
By Taylor expanding, we see that $F(\theta) \approx e^{i A_w \theta}$. This is not algebraic trickery—it is a geometrically necessary phenomenon. The initial overlap $\braket{\phi}{\psi}$ is our reference scale. As the state evolves, its projection onto $\bra{\phi}$ rapidly changes phase if $\ket{\psi}$ and $\ket{\phi}$ are nearly orthogonal.

This phase growth rate—the derivative of $\arg F(\theta)$ at $\theta = 0$—is $\text{Re}(A_w)$. A large real weak value thus leads to local phase inflation. The geometry forces $F(\theta)$ to behave like $e^{i \Omega \theta}$ with $\Omega = \text{Re}(A_w) > \Lambda$, hence superoscillation. This phenomenon arises because the space of projections from $\ket{\psi(\theta)}$ onto $\bra{\phi}$ is curved, and the weak value measures how $A$ tilts that curvature.

\subsection{Worked Derivation: From Weak Values to superoscillation}

We now explicitly derive the connection between weak values and super-oscillatory behavior, without reference to measurement devices. Let \(A\) be a Hermitian operator with spectrum in \([-\Lambda, \Lambda]\), and define

\[
F(\theta) := \frac{\bra{\phi} e^{i\theta A} \ket{\psi}}{\braket{\phi}{\psi}}
\]

We assume \(\braket{\phi}{\psi} \neq 0\), though possibly small.

Expanding the numerator in Taylor series:
\begin{align*}
F(\theta) &= \frac{1}{\braket{\phi}{\psi}} \sum_{n=0}^\infty \frac{(i\theta)^n}{n!} \matrixel{\phi}{A^n}{\psi} \\
&= 1 + i\theta A_w - \frac{\theta^2}{2}(A^2)_w + \dots
\end{align*}
where \((A^n)_w := \frac{\matrixel{\phi}{A^n}{\psi}}{\braket{\phi}{\psi}}\). If \(|\theta|\) is small and higher-order terms are negligible, we approximate:
\[
F(\theta) \approx e^{i A_w \theta}.
\]
This suggests that \(F(\theta)\) oscillates locally with frequency \(\text{Re}(A_w)\).

However, this frequency can exceed \(\Lambda\), the global spectral bound of \(A\). How is this possible?

\paragraph{Geometric Explanation.} The numerator tracks the deformation induced by \(A\), while the denominator sets the reference geometry. If \(\braket{\phi}{\psi} \ll 1\), even a mild deformation becomes magnified. The lens formed by \(\bra{\phi}\) “sees” a rapid change in phase due to this curvature. Thus, \textit{the apparent high frequency is a geometric inflation of local phase}, not a violation of the spectral bounds.

This is the essence of a superoscillation: a slow function with fast derivatives, induced by a geometric projection.

\paragraph{Bargmann Triangles and Curvature.}
The argument of the weak value corresponds to a triangular loop in projective space whose geometric phase is measured by the Bargmann invariant \footnote{The Bargmann invariant captures the geometric phase accumulated around a triangle formed by three states in projective Hilbert space, linking interference and curvature. The connection between weak values and geodesic triangles in projective Hilbert space has recently been emphasized in \cite{Pan2024}, where the Bargmann invariant is used to derive Pancharatnam phases and interpret weak values geometrically. Our approach is complementary: we treat the Bargmann triangle as part of a larger deformation mechanism, where weak values emerge not only as geometric phases but as amplifications induced by a warped sesquilinear structure. 
}:
\[\Delta(\psi, \psi_\theta, \phi) = \braket{\psi}{\psi_\theta} \braket{\psi_\theta}{\phi} \braket{\phi}{\psi}.\]
This captures how curvature affects the total phase — the lens effect being a geometric twist in state space.

\section{Conclusion and Future Work}
We have framed the weak value as a \textbf{ratio of geometric deformation}. It is a generalized Rayleigh quotient whose meromorphic structure on projective space dictates the existence of "anomalous" values, which, in turn, generate superoscillations. The power of this framework is conceptual and organizational, providing an intuitive language for designing systems that exploit these effects in high-precision metrology and super-resolution imaging \cite{Potocek2017}.

The same formalism naturally extends to a more general setting: given two observables $A_{1}$ and $A_{2}$, one may consider the expression 
\[
W_{A_1, A_2} = \frac{\matrixel{\phi}{A_{1}}{\psi}}{\matrixel{\phi}{A_{2}}{\psi}},
\]
which compares the geometric deformation induced by $A_{1}$ relative to that of $A_{2}$. The familiar weak value corresponds to the special case $A_{2}=I$. We intend to explore this direction in future work \cite{MannucciForthcoming}.
\subsection{Beyond Identity: The Ratio of Two Geometric Deformations}

The weak value formula is often written with the identity operator as reference:
\[
A_w = \frac{\matrixel{\phi}{A}{\psi}}{\braket{\phi}{\psi}} = \frac{\matrixel{\phi}{A}{\psi}}{\matrixel{\phi}{I}{\psi}}.
\]
But this is a special case of a more general expression involving two operators:
\[
W_{A_1, A_2} := \frac{\matrixel{\phi}{A_1}{\psi}}{\matrixel{\phi}{A_2}{\psi}}.
\]
This expression compares the deformation induced by \(A_1\) with that induced by \(A_2\), as seen through the same initial and final states. When \(A_2 = I\), we recover the usual weak value.

Geometrically, this represents a ratio between two sesquilinear forms:
\[
W_{A_1, A_2} = \frac{\matrixel{\phi}{A_1}{\psi}}{\matrixel{\phi}{A_2}{\psi}} = \frac{g_{A_1}(\phi, \psi)}{g_{A_2}(\phi, \psi)},
\]
which compares how two different observables deform the geometry of Hilbert space along the same trajectory from $\ket{\psi}$ to $\bra{\phi}$. This generalizes the weak value (where $A_2 = I$) to a broader notion of geometric amplification.

which captures how the geometry induced by \(A_1\) differs from that of \(A_2\), relative to the same pair of states. This is the foundation for a more general geometric theory of “weak ratios,” where both numerator and denominator carry structural information.

This perspective opens a promising path: different observables may correspond to different geometric distortions, and their relative effect encodes more than either alone. A full theory of such ratios is the subject of a forthcoming article \cite{MannucciForthcoming}.
\section{Future Directions and Operational Amplification}

This work has shown that weak values emerge naturally from a geometric deformation of Hilbert space, and that this deformation predicts superoscillatory behavior. Several important directions remain open for further development.

\subsection*{Geometric Amplification in Quantum Circuits}

A key next step is to investigate how the geometric dilation effect can be leveraged in quantum computing architectures. Weak value amplification—especially in composite systems—offers a potential strategy for boosting signal features without adding external energy. Such dilation could serve as a native amplification mechanism in quantum circuits, where small gate-induced deformations are magnified via coherent weak lensing. This concept might be particularly impactful for:

\begin{itemize}
  \item enhanced sensitivity in interferometric subroutines,
  \item dynamic re-routing via geometric phase control,
  \item sub-threshold signal detection in noisy circuits.
\end{itemize}

We intend to model and simulate such circuits in follow-up work.

\subsection*{The Weak Derivative as Local Generator}

An important mathematical structure suggested by this work is the interpretation of the weak value as a complex directional derivative:
\[
\left.\frac{d}{d\theta} \arg \left( \braket{\phi}{e^{i\theta A} \psi} \right) \right|_{\theta=0} = \text{Re}(A_w).
\]
This formulation justifies calling the weak value a local generator of phase inflation. It acts not only as a diagnostic but as a geometrically meaningful object encoding the rate of deformation relative to the post-selection frame.

\subsection*{Weak Geometry Beyond Two States}

While this article focused on the two-state weak framework, the meromorphic nature of the weak value function on $\mathbb{CP}^n \times \mathbb{CP}^{n*}$ suggests deeper geometric structures. We conjecture that weak lensing could be recast in terms of a complex line bundle with singularities, and that the study of these singularities may shed light on topological transitions and coherence plateaus in quantum dynamics.

\section*{Acknowledgements}
The geometric insight that anchors this work was born in 2006, during a remarkable seminar by Yakir Aharonov on the foundations of weak measurement. I am deeply grateful to Daniele C. Struppa for connecting me to Jeff Tollaksen, whose invitation to that lecture proved transformative. Upon seeing the weak value formula for the first time, I immediately perceived it not as a mere measurement outcome, but as a sign of geometric deformation—a lens reshaping the very fabric of quantum state space. This paper attempts to honor that moment by making its implicit geometry explicit.

\appendix
\appendix
\section*{Appendix: Computational Simulation of superoscillations}
This appendix reproduces the geometric amplification and superoscillatory behavior arising from weak values, using Python with NumPy, SciPy, and Matplotlib.

\subsection*{Part 1: Weak Value from Geometric Deformation}
We construct a weak measurement scenario for a single qubit using the observable \( A = \frac{1}{2}\sigma_z \). The weak value is computed from the deformation of the initial state by the observable and the post-selection:

\begin{verbatim}
import numpy as np

a_param = 3.0  # Strength of post-selection lens

# Define |+x> and |-x>
plus_x = (1 / np.sqrt(2)) * np.array([1, 1])
minus_x = (1 / np.sqrt(2)) * np.array([1, -1])

# Pre- and post-selected states
psi = plus_x
phi = (plus_x + a_param * minus_x) / np.sqrt(1 + a_param**2)

# Observable: sigma_z / 2
A = 0.5 * np.array([[1, 0], [0, -1]])
A_psi = A @ psi

# Projections and weak value
proj_orig = np.vdot(phi, psi)
proj_deform = np.vdot(phi, A_psi)
A_w = proj_deform / proj_orig

print(f"Weak value A_w = {A_w:.4f} with magnitude |A_w| = {abs(A_w):.2f}")
\end{verbatim}

This shows that the weak value acts as a geometric magnifier: even mild deformations can result in large effects if the pre- and post-selection are nearly orthogonal.

\subsection*{Part 2: superoscillations in Multi-Qubit Systems}
We now simulate a system of \( N = 20 \) qubits. The geometric amplification from the weak value yields a locally oscillating signal with frequency far exceeding the operator's bandlimit:

\begin{verbatim}
import matplotlib.pyplot as plt
from scipy.linalg import expm

N = 20  # Number of qubits
A_w_N = N * A_w
Lambda = N / 2  # Bandlimit of full observable
theta = np.linspace(-0.2, 0.2, 400)

# F(theta): actual evolution
F_theta = ((phi.conj() @ expm(1j * theta[:, None, None] * A) @ psi)
           / (phi.conj() @ psi))**N

# G(theta): ideal prediction from geometric lens
G_theta = np.exp(1j * A_w_N * theta)

plt.figure(figsize=(12, 6))

plt.plot(theta, np.real(F_theta), label='Re[F(theta)] (Actual)', lw=3)
plt.plot(theta, np.real(G_theta), '--', label='Re[e^{i A_w theta}] (Predicted)', lw=2)
plt.xlabel("theta")
plt.ylabel("Real Part")
plt.title("superoscillations via Weak Value Amplification")
plt.legend()
plt.grid(True)
plt.tight_layout()
plt.show()
\end{verbatim}

The match between the predicted exponential \( e^{i A_w \theta} \) and the actual evolution demonstrates that superoscillations are not an artifact of measurement, but a geometric necessity resulting from the weak value mechanism.
\subsection*{Part Three: Code Repository.} A GitHub repository with all code and simulations illustrating the geometric viewpoint and superoscillatory amplification is maintained at: \\
\url{https://github.com/Mircus/quantum-weak-values-geometry}.


\begin{thebibliography}{99}
\bibitem{Aharonov1988} 
Y. Aharonov, D. Z. Albert, and L. Vaidman, "How the result of a measurement of a component of the spin of a spin-1/2 particle can turn out to be 100," \emph{Physical Review Letters}, vol. 60, no. 14, p. 1351, 1988.

\bibitem{Berry1994} 
M. V. Berry, "Faster than Fourier," in \emph{Quantum Coherence and Reality}, eds. J. Anandan and J. L. Safko, World Scientific, 1994, pp. 55-65.

\bibitem{Kempf2000} 
A. Kempf, "Black holes with quantum hair," \emph{Physical Review D}, vol. 61, no. 8, p. 081502, 2000.

\bibitem{Berry2006} 
M. V. Berry and S. Popescu, "Evolution of quantum superoscillations and optical superresolution without evanescent waves," \emph{Journal of Physics A: Mathematical and General}, vol. 39, no. 22, p. 6965, 2006.

\bibitem{Dressel2014} 
J. Dressel, M. Malik, F. M. Miatto, A. N. Jordan, and R. W. Boyd, "Colloquium: Understanding quantum weak values: Basics and applications," \emph{Reviews of Modern Physics}, vol. 86, no. 1, p. 307, 2014.

\bibitem{Colombo2011} 
F. Colombo, I. Sabadini, D. C. Struppa, J. Tollaksen, and Y. Aharonov, "Some Mathematical Properties of superoscillations," \emph{Journal of Physics A: Mathematical and Theoretical}, vol. 44, no. 36, p. 365304, 2011.

\bibitem{Bargmann1964} 
V. Bargmann, "Note on Wigner's Theorem on Symmetry Operations," \emph{Journal of Mathematical Physics}, vol. 5, no. 7, pp. 862–868, 1964.

\bibitem{Potocek2017} 
V. Potoček, F. M. Miatto, M. V. Berry, N. Brunner, and S. G. Popescu, "Quantum-limited superresolution of an incoherent source," \emph{Physical Review Letters}, vol. 119, no. 25, p. 250501, 2017.

\bibitem{Sabadini2019} 
I. Sabadini and D. C. Struppa, \emph{superoscillations: Theory, Applications, and Experiments}, Springer, 2019.

\bibitem{Dressel2012} 
J. Dressel and A. N. Jordan, "Sufficient conditions for uniqueness of the weak value," \emph{Journal of Physics A: Mathematical and Theoretical}, vol. 45, no. 1, p. 015304, 2012.
\bibitem{Mannucci2025}
M. A. Mannucci, ``When the Weak Becomes Strong: Effective Observables via Time-Symmetric Quantum Selection,'' \emph{arXiv preprint} arXiv:2507, 2025. Available at: \url{https://arxiv.org/abs/2507.09716}.
\bibitem{MannucciForthcoming}
M. A. Mannucci, ``Generalized Weak Values,'' forthcoming manuscript, in preparation (2025).
\bibitem{Pan2024} Y. Pan and M. L. Ge, ``Weak Values and Pancharatnam Phases on Geodesic Triangles in the Complex Projective Space,'' \emph{arXiv preprint arXiv:2506.17132v2}, 2024.
\end{thebibliography}
\end{document}